\documentstyle[epsfig,12pt]{article}
\begin{document}

\begin{center}
{\Large \bf IS THERE CHAOS IN THE  ELECTRON MICROSCOPE?}
\vspace*{0.5cm} \\
D.Kunstelj \\
{\em Department of Physics, Faculty of Science (PMF), University of
Zagreb, Bijeni\v cka c.32,10000 Zagreb, Croatia} \\
\vspace*{0.3cm}
M.Martinis and A.Kne\v zevi\' c \\
{\em "Rudjer Bo\v skovi\' c" Institute, Bijeni\v cka c. 54,10000
Zagreb, Croatia}

\vspace*{0.5cm}
                        {\bf Abstract}
\end{center}

High-temperature superconductor $Bi_{2}Sr_{2}CaCu_{2}O_{8}+x$ has been
investigated by the high-resolution electron microscope (HREM) and the
photodensitometric technique  in order to resolve the sub-atomic shifts
in the modulated structure.
This investigation has shown that the background noise is
pronounced, indicating some kind of ordering in certain
crystallographic directions.
We have found that the "mapping" of the charge densities by the
inverse Fourier transform of the diffracted electron image and the
connections to the crystal structure can be easily established in the
case of elastically, as well as the inelastically scattered electrons.
The order of the details in so obtained "inelastic" electron
"micrographs", as well as the dependence of the structure of the EM image
on the initial conditions of the inverse Fourier transform, leads to the
conclusion that there is an order emerging from the  diffuse scattered
electron diffraction patterns.
Data sets from digitalized densitograms are analysed by the Rescaled range (R/S) method
 in order to study fractal and long-range correlation properties of
the inelastic scattered electrons in the "noise" region between Bragg
reflections.
The results of the R/S analysis showed that the value of the Hurst
exponent was H$\neq$ 0.5, indicating that the "ordering" of the noise
in the Inverse (Fourier) space was much more pronounced at
longer distances (about 440 atom spacings),
then for shorter distances of  about 55 atom spacings.

\section{Introduction}

Electron micrographs of high resolution (HREM) show an considerable
amount of the background noise superimposed on regularly arranged network
of motifs representing the order in the crystal lattice. The background
noise has various origins. One is the inelastic
interaction of incoming electrons with host electrons and phonons of the sample under
investigation.
The final image which is the convolution of these inelastic, and other
effects will be disordered to some extent,
showing general regular crystal motif smeared by diffuse scattered
electrons. Now, we may ask the question: Are those diffusely scattered
electrons really stohasticaly distributed in space
as seen in Fig.1a,b and Fig.2b,c, or there is some kind of the order
emerging from the interaction of the EM electrons with the charge
distribution in the specimen? Is there really some regularity
in the diffuse scattering of  the EM electrons transmitting the sample
("diffuse background"). In other word, do we see some chaotic order
in high resolution electron micrographs?

In order to study the contribution of the diffusely scattered
electrons to the micrograph, we should be able to separate their
effect from the effect of elastically scattered electrons on  the
intensity of the electron beams impinging the photographic plate
in the electron microscope. For this purpose we process the EM
micrograph following the events of the image formation in the
microscope. Apart from the modulation of the EM electron
intensities by the instrument itself (transfer function) and the
mode of the operation conditions applied in detection of the
particular motif under investigation (focussing conditions),
formation of the image consists of diffraction of the EM
electrons, their interference and the amplitude reconstruction.
This processes are well described by the Fourier transforms and
may be simulated by a suitable computer program. So the EM image
may be processed following the events in the electron microscope.
The computer program should be able to separete elastically as
well as inelastically diffracted "electron beams" which contribute
to the formation of the image. One of them is ProFFT (Professional
Fast Fourier Transform), which enables us to explore the Brillouin
zones of the crystal one by one, their sections, and to map the
results of the interactions of the EM electrons with the charge
densities of the host crystal.

\section{Experimental}

\subsection{Basic electron microscopy of the $Bi_{2}Sr_{2}CaCu_{2}O_{8}+x$
non-modulated samples}

The figure of  non-modulated structure of
$Bi_{2}Sr_{2}CaCu_{2}O_{8}+x$ high-temperature superconductor
(T$_{sc} = $ 85 K), obtained by HREM [1,Fig.2] was scanned and
entered to ProFFT program. This image (reproduced with the laser
printer) shown in Fig.1a, resembles very well the original. White
dots in the rectangular array reperesent dominantly the Bi atom
columns, which are the strongest electron scatterers, the {\bf c}
projection of the elemental cell is shown on the right side {\bf
(a,b)} of the figure. The laser aided optical Fourier transform
(OFT) which was made using the EM micrograph negative [1], is
shown in the same figure, with the main reflecions labeled as the
insert down left. The diffraction pattern of this figure, i.e.,
its Fourier transform is obtained by applying the ProFFT and
presented in Fig.1b. The main reflections of the unit cell are
labelled, the "filtering ring" and the "filtering cross" are shown
as well. The "filtering ring" allows the diffusely scattered
electrons with ${\bf k}(x')^{2}+{\bf k}(y')^{2}={\bf
k}(x',y')^{2}=const.$ ($x'$ and $y'$ being the coordinates of the
Inverse space), i.e. with a certain kinetic energy, to enter the
next step, the Inverse Fast Fourier transform (IFFT), of the image
formation. This way, we obtain the picture, or the map, of  the
diffuse scattered EM electrons which suffered a certain amount of
the energy gain or loss when passing the sample. If we use "the
cross filtering", we can choose any ${\bf k}(x')$ and/or ${\bf
k}(y')$, their interval, or their linear combination in the
inverse (Fourier) space (${\bf k}(x')+{\bf k}(y')={\bf
k}(x',y')$), and to show the map of the energy losses, or gains,
of the transmitting electrons in specific directons, and to see if
this losses are Stohastic or Chaotic in their nature. Fig.2a,[1],
shows nearly sinusoidal behaviour of the contrast along the rows a
and b of the atoms (a), level fluctuations is caused by the uneven
blackening of the negative of the EM photograph. The Fourier
transform of the digitalized trace shown in Fig.2a is presented in
Fig.2b, the peak correspons to the (200) reflection, with the
crystal plane spacings of 2.715 A.U. The same was done for the b
direction, and the result is shown in Fig.2c. To the left (I. and
the II. Brillouin zones.) and to the right (higher Brillouin
zones.) of labelled Bragg reflections, the photo-densitometric
trace of the diffuse background shows no distinctive regularities,
and, it this step of the observations may  be described as purely
stohastic, or nondeterministic. Subsidiary maxima close to the
(020) reflection in Fig.2c, may be due to the initialization of
some crystal ordering [1], this matter is elaborated in Ref.1, and
is beyond the scope of the present paper.

 Now, we can do the procedure as described above, i.e., to select a
certain combination of ${\bf k}(x')$ and ${\bf k}(y')$, to make an
IFFT procedure, and to compare the density map so obtained with
the crystal structure. In order to check the whole procedure the
crystal structure itself was "purified" by blocking the background
between the Bragg reflections, once (Fig.3a) and repeating the
procedure, i.e., twice (Fig.3b). It can be seen that the contrast
and the informations about the details of the figure are augmented
this way. This is much better shown in Fig.4a (enlarged detail of
Fig.1a), and  4b (enlarged detail of Fig.3b), where, for the
precise measurements of the charge density positions, the sketch
of the crystal lattice is superimposed upon the EM image (White
arrows on the left are the orientation marks). It is well seen
that the main charge densities, i.e., the positions of atoms (atom
columns) are centered at the crystal lattice nodes, the lattice
parameters in the a and b directions are nearly the same, being
about 5.4 A.U.[1]. So, this figure, which we interpret as the
projecion of the charge densities of the crystal lattice, can
serve as the standard for the further investigations.

\subsection{Experimental situation and some theoretical considerations}

First we do short overlook of the experimental situation in our case, and
 then some theoretical considerations and approximate calculations.
Let us see what is going on with the 200 keV  electrons when
passing throught the crystalline sample, 10 to 20 nm thick, at the
approximate temperature of  about 400 K, with the relativistic
speed of  about 2/3 of the speed of light. They "see" the crystal
lattice, which vibrates according to  $k_{B}T=hf$ , with the
frequency of the order of 10$^{13}$ Hz, $k_{B}$ being the
Boltzmann constant, and $h$ the Planck constant. The "time of
flight" of the 200 keV electron through the sample of  about 10 nm
is simply then about half of  10$^{-16}$ seconds, which is 5/10000
of the estimated period of the period of  atom vibrations. So the
every wavefront of the EM electrons see  "frozen"  lattice, but,
each incoming wavefront in a different state of  the lattice
vibrations, longitudinal and transverzal. As the exposure of an EM
negative takes one to two seconds, we get the picture of about
$n=t(exposure)/t(flight)=4\cdot 10^{16}$ slightly different states
of the atom positions, vibrating inside the "volume" of an atom,
with the phonon spectrum given by the crystal lattice and its
atoms, being at the temperature of about 400 K. So the EM
electrons "see" the phonon spectrum of the sample, that is all
vibration states, and "transfer" it to the EM negative. The
ergodicity of the system is thus fulfilled, and so the Liouville's
theorem holds.

The average positions of the atoms thus give spread (Debye-Waller
broadening) Bragg reflections, i.e., the Fourier transform of the
lattice. The inelastic interactions of the EM electrons with
phonons may give what we usually call the background noise, which
may be of the chaotic nature,
 that is, with some order dictated by the phonon spectrum and the host
electron states.   The next event is the interaction of the EM
electrons with the host electrons; that bounded within the inner
electronic shells of the atoms, the valency electrons, and the
loosely bonded "wandering", i.e. conduction electrons. The
interactions of the EM electrons with the inner shells
(K,L,M,$\ldots$), with the energy exchange of the order of  keV-s,
shall not be considered here, because this matter is covered by
the very well known EELS (Electron Energy Loss Spectroscopy), and
the diffracted inelastically scattered electrons lie far away the
usually recorded diffraction pattern, that is, far from the origin
of the Fourier transform of the observed crystal lattice. Now, let
us see what we can really do by making some short and simplified
calculations of  the inelastic interactions
 of the EM electrons and the loosely bounded electrons in our sample. The
energy loss calculation may be simply argued by the observation of
the background diffuse scattering within the 1st and the 2nd
Brillouin zones in the diffraction pattern (Fourier transform) of
the crystal (Fig.1a., insert down left, Fig.1b.), which is
obtained by making the Fourier transform of the crystal lattice
using the ProFFT (Professional Fast Fourier Transform) program.
The calculations are based on the use of the Bragg equation, which
is extended with the inelastic member, i.e., in the vector form
this equation is now:  $\bf k - k_{0} =  g + dk$ ; within the 1st
and the 2nd Brillouin zones $\bf g = 0$, and the $\bf dk$ is the
inelastic member which can be calculated from its position in the
1st and the 2nd Brillouin zones using the diffracion pattern as
follows. Let dk coming from the the energy loss being exactly at
the half of the (200) position, that is at the edge of the 1st
Brillouin zone (1st.B.z.) in the [100] direction. Then $dk = g/2 =
1/2d(200)$; with $d(200)$ in our case being 2,715 AU = 2,7$\cdot
10^{-10}$ m, so $dk = 1,84\cdot 10^{9} m^{-1}$. The energy loss we
calculate from the expression for the kinetic
 energy of a free particle:  $E=(hk)^{2}/2m$, h=6,6236$\cdot 10^{-34}$ Js is the
Planck constant, m=9,11$\cdot 10^{-31}$ kg, the mass of an
electron, $k=dk$ for the edge of the 1st.B.z., so the energy loss
E = 8,15$\cdot 10^{-19}$ J = 5 eV. The energy is proportional to
k$^{2}$, so the energy loss of an inelastically scattered electron
at the edge od the 2nd.B.z. ($k_{2} = 2k_{1}$)  is of the order of
20 eV. So, by the simple selection of the part of  the Inverse
Space generated by the Fourier transform (Paragraph 1.) we can
investigate the the electron energy loss spectra and the space
distribution of inelastically scattered electrons in the range up
to 20 eV by using the FFT and IFFT procedures of the ProFFT
program as described above.

\subsection{The Rescaled Range (R/S) analysis of  the "noise" in the
densitograms of  the High Resolution Electron micrographs}

Data sets of 512 (2$^{9}$) points (about 55 atom spacings) are cut
from the whole of 4096 (2$^{12}$) points (about 440 atom
spacings), from digitalized densitograms [1] of  ten different
traces along a[200] and b[020] directions of the BISCO
(Bi2Sr2CaCu2O8+x) crystal, in order to analyse the existence of
short/ or long range correlations in the "noise" region between
Bragg reflections. An example of densitogram [1] is shown in
Fig.2a.

As we believe that such "noise" arises from interactions of
incoming electron beam with crystal lattice, such as
electron-phonon and electron-electron inelastic scatterings, it
could bring some informations from the system. In terms of
nonlinear analysis we expect for signals to have a long-range
correlation properties and not be white noise (complete random
data).

The Rescaled range (R/S) method of nonlinear analysis was used to
study the correlation properties in our data sets [2,3,4]. The
method is related to the evaluation of the Hurst scaling exponent,
H. Different values of the Hurst exponents correspond to different
correlation properties.

For the signal represented by the data set u(n), n = 1,....., N in
one crystalographic direction we calculated the running average
$\overline u(n)$ and the accumulated deviations from the average
$X(l,n)$;

\[ {\overline u(n)} =\frac{1}{n} \sum_{k=1}^{n} u(k) \]
\[ X(l,n)=\sum_{k=1}^{l} [u(k)-{\overline u(n)}] \]

The quantity called the range R(n) of X(l,n) and the standard
deviation S(n) are defined as follows:

\[ R(n)=max_l X(l,n)-min_l X(l,n) \]
\[ S(n)=\sqrt{\frac{1}{n} \sum_{k=1}^{n} (u(k)-{\overline u(n)})^{2}} \]

The "rescaled range" is defined as a ratio between R and S (R/S).
The power law scaling according to Hurst is:

 \[ R(n)/S(n)\sim n^{H} \]
where H is the Hurst exponent. If a signal represents white noise
(uncorelated signal) then H=0.5. If H$\neq$0.5, the long-range
correlations (memory) are in the system.

The scaling exponent H is evaluated from the log(R/S) vs. log(n)
plot using the least square fit procedure.

The running procedure
is as follows: we calculate R/S for one box of n points, starting with first
two points. In the next step we add one point more and calculate R/S
for the wider box and so on, until the box has length N of starting data
set. In this way we always hold one box with fixed left and  moving right
 end.

The results of the analysis are shown in Fig.5. This figure is the
 representative of  21 analysis worked on traces of the
 BISCO crystal lattice.

\subsection{The ProFFT analysis of the diffuse scattering  in HREM micrographs
of BISCO}

The evolution of an physical system is usually described in the 6D
phase space, which is defined by 6 degrees of  freedom; 3 space
coordinates and 3 impulse coordinates ({\bf r},{\bf p}). A priori
all the states of the system are equally probable (Liouville's
theorem) and the probability of the state of the system is equal
to the time average of  the ensamble (Ergodic theorem; ensamble
may be one particle). The "volume" of the phase space is
determined with the total energy of the system. The phase space
can be "divided" (separated) into subspaces according to the
specific problems of the system investigated. So, we may choose to
represent the system in Cartesian coordinates ({\bf r}(x,y,z)), so
called the Direct space; or/and we may use the impulse coordinates
({\bf p}(p$_{x'}$,p$_{y'}$,p$_{z'}$)), so called the Inverse
space. This two spaces are then connected with the Fourier
transformations. The  trajectories of of the system which
undergoes the chaotic behaviour in the 6D phase space are usually
presented as its projections known as the Poincare sections, in
order to get the 2D presentation in, say, (x,p$_{x'}$) projection.
We may also choose another coordinate subsystems of the 6D phase
space to present our problem, so we can take the Direct space
(charge distribution in the crystal) and the Inverse space (i.e.
the diffraction pattern in the reciprocal space), which two are
then connected by the Fourier transformations.

The Direct and the Inverse space are easy to explore experimentaly
by the use of the High Resolution Electron Microscopy (HREM),
taking the image and the corresponding diffraction pattern of the
crystal. The HREM data can be digitalized and further explored by
the computer programs which are based on the use of the Fourier
transforms (ProFFT). In this respect, following our R/S analysis (
paragraph 2.3), we first explored the situations in $x$ and $y$
directions (HREM micrographs), and the k$_{x'}$ and k$_{y'}$
directions (the diffraction patterns). Here we concentrated on the
diffuse scattered electrons in the 1st and 2nd Brillouin zones,
that is from the origin up to (200) reflections (Fig.1b). If we
take the horizontal part of the cross, which we obtain by proper
masking of the whole diffraction pattern ( the [020] and the
opposite direction, down left insert in Fig.1b), and the use of
ProFFT programm to perform the Fourier transformation, we obtain
the situation as shown in Fig.6a. Analogously, taking the vertical
part of the cross, which contain the diffuse scattered electrons
in the [200] direction, to contribute to the image formation, we
obtain the distribution of the diffuse scattered EM electrons with
an energy loss, as shown in Fig.6b. In both figures, Fig.6a and b,
we can recognize white bands ("stripes") on dark background and
dark bands ("stripes") on the pale (white) background. The stripes
are perpendicular to the direction of the analogous reflection in
the diffraction pattern, they are approximately equidistant, and
more or less "paired". They can also "merge" to each other, which
is somewhat better seen in Fig. 6a. The white and the dark stripes
are nearly of the distance as the atom
 columns in Fig.1a. So we can interpret them as the result of the inelastic
scattering of the EM electrons at the atom sites, with the absorption
 of some energy (dark stripes on white bands), and the re-emission of
similar energy (white stripes on the dark bands).
%--------------------------------------------------------------------
%    (Tu se ja pitam, pa i tebe: da li je ta interpretacija u redu?)
%--------------------------------------------------------------------

This figures do
 not say anything about the possible charge distribution and the inelastic
scattering between the atom columns positions, which may be present if
 there is some charge transfer between the neighbouring atoms. To explore
 this possibility we may use the whole cross from Fig.1b, or the half of
 it, that is from the origin of the reciprocal space to the half distance
 to the (200)  and the (020) maxima. This way we are taking only the
inelastically scattered electrons with the energy losses in the 1st Brillouin
 zone. We can easily calculate this band of energies (The basis of this
calculation is already given in 2.2.). Moreover, we can mask any part of
the diffraction pattern, and by the use of the ProFFT, see the
distribution of the diffuselly scattered electrons with certain energy loss.
(Here we emphasize the regularity of the order of the stripes showing
the regularity of the energy losses of diffusely scattered electrons,
in relation to the diffuse background in the diffraction patterns
(Fig.1, Fig.2b,c),
which are "chaotic". This regularity may be explained with the
quantization of the energy transfer between the EM electrons and the
charge and phonons in the sample, as well as with the specific atom
positions in the crystal lattice.)
Thus, in Fig.7a and 7b we present the images of inelastically scattered
electrons in the 1st and the 2nd B.zones superimposed
 on the crystal grid, which was taken from Fig.1a in order to correlate
them to the positions of the atom columns ({\bf a} and {\bf b} are
as in Fig.1a, white arrows on the left-down of the images in
Fig.7a and 7b are  reference marks.). The diffusely scattered
electtrons are positioned at the atom sites, i.e., on the
crossings of superimposed crystal grid, and between them.
 The patches are "dark" on the white background, this shows the
absorption in this directions, and "white" on dark background, which shows
 the diffraction in that direction. This "secondary" diffractions are
 probably due to the elastic diffraction of the EM  electrons which were
already subjected to energy loss by the inelastic scattering. This
two possibilities are, for huge number of the EM electrons
involved in the image formation, the same, so the number of
"white" and "dark" patches is expected to be equal, as easily seen
by the inspection of Fig.7. In addition, white (pale) and the dark
(gray) "bridges" between inelastically scattered electron maxima
we interpret as inelastic interactions with the valence electrons
and the phonons in the {\bf a} and the {\bf b} directions,
 i.e. along the directions of the elemental cell. Also, it has to be noted
that the diffusely scattered maxima ("white" as well as "dark") are
equally positioned between the crystal nodes, which may be the result of
a very different causes that has to be still more explored. For example,
"white" as well as the "dark" patches may be the result of
  different, but quantizied, energy losses which are close in the amount
and the scattering direction. Or, they can be the result of the extended
crystal defect, which is not recognized in the direct EM image (Fig.1a).

However, the regularity of the diffuse scattering is evident. This phenomenon
is demonstrated by the experiment in which the initial conditions are
slightly changed, i.e., the cross, which defines the number of inelastically
scattered electrons which form the image, can be slightly changed,
say by narrowing it, as was done for Fig.7b in comparison to Fig.7a. The
general appearance of this figures is nearly the same, they only differ in
details, for example in the intensity, but not the position of the "white"
and the "dark" patches. This was to be expected, because the
nature of the physical processes was not changed by tis intervention,
only the number of the inelasticaly scattered electrons which form the image.
 This is the result of the chaotic behaviour of the system in the
case we do not go far in change from the initial conditions, and if the
Liapunov coefficient is not to big for the interval of the exposition time.
Fig.7a and Fig.7b show that our presumptions are fulfilled.

In our investigations of  the image formation in the Electron
microscope by exploring the inelastically scattered electrons
contribution we used a non-standard approach in the analysis of
the chaotic nature of the system by observing the connecions of
the Direct and the Inverse space by  applying the Fourier
transformations.(Usually, the possible chaotic nature of the
physical system is investigated by the analysis of the {\bf
r},{\bf p}-pairs projections of the Phase space (Poincare
sections). Here, our "Poincare sections"
 of the Phase space are {\bf x} and {\bf y} coordinates in the Direct, crystal space,
which are connected to the analogous {\bf p}$_{x}$ and {\bf
p}$_{y}$ coordinates in the Inverse space by the Fourier
transformations).

\section{Disussion}

We show that in the Direct  space of the crystal there is a
definite order of atoms (Fig.1a, Fig.2a, Fig.3a,b, Fig.4a,b).Also,
the main result of our analysis of the inelastically (diffusely)
scattered electrons (Fig.6 a,b, Fig.7a,b) is that there is the
definite order in their diffraction, which was demonstrated by the
use of the Fourier transform as described in the previous section.
The independent R/S analysis (Fig.5) also showed that there is a
long range order in, at the first sight stohastically chaotic
background in the Inverse space (diffuse background in Fig.1b and
Fig.2b,c). So we concluded that our process of the image formation
in the Electron microscope is chaotic in its nature. The physical
background of  such conclusion is proposed in the paragraph 2.2 of
the paper.

To
 illustrate this standpoint we enclose one more result of our investigations
in Fig.8. The results presented in the previous paragraph and the
considerations of the physical system we explored allow us to propose the
 model for the theoretical elaboration of the problem, which task is far
from the matter here elaborated. First, the possibility of  investigation
 of  the influence of  crystal oscilations on the electron-phonon energy
exchange  we see in connection to the wery well known model of  the
situation known as The Fermi acceleration [5]. In this model, as shown in Fig.
8a, the particle (our EM electron) meets the potential, which oscillates
with some amplitude (a) with a certain frequency. The amplitude and
 the frequency may be estimated from the Debye-Waller factor for the system
under investigation, if  the temperature of the sample is known.
In the next step this model may be connected to the Quantum
billiard system [6], in which the billiard walls oscillate to
produce the Fermi acceleration, thus simulating the oscillating
crystal lattice. The result may be as in Fig.8b, in which a
contours of the quantum mechanical wave function, whose classical
counterpart has chaotic behaviour, are shown [6]. In this sketch
of the wave function solid lines indicate positive values, and the
dashed lines negative values, which may persent the emission and
the absorption of inelastically scattered electrons respectively.
This figure
 may be connected with our observations, one example is shown in Fig.8c,
where the "positive" parts are shown as white areas containing somewhat
darker patches, and the "negative" parts are shown as dark areas with a
pale patches; also presenting the absorption and re-emission processes,
as proposed in the previous paragraph. Figure 8c was obtained by the
 same procedure as applied for Fig.7, but with electrons filtered by a
ring (Fig.1b, insert up right), thus allowing the inelastically scattered
electrons of approximately same energy, but in an energy band deffined by
 the wideness of the ring, to enter the Fourier transform procedure (ProFFT),
and to show the inelastically scattered electron density map.

The situation
with "white" and "dark" patches is similar to that in Fig.
2.7. They are of approximately same size, distributed at approximately
same distance, somewhere connected with the "bridges", but less
 than in Fig.7. Thus we conclude that the diffusely scattered electrons
with approximately same lost or regained energy (Fig.8c) are more localized
than diffusely scattered electrons within an wider energy band (1st and
2nd  Brillouin zone, Fig.7), the result which was expected. The connections
of  the patches showing the positions of the diffusely scattered electrons
may show the ways which may be used by the conduction electrons when
conducting the electric current in the "normal" state and in the
superconducting state as well.
\vspace*{0.5cm} \\
ACKNOWLEDGEMENT

The authors acknowledge the use of  the data from Ref.1. This work
was supported by the Ministry of Science and Technology of Croatia
under the contract 119203 and 0098004.

FIGURE CAPTIONS

\begin{description}

%\begin{itemize}
\begin{itemize}

\item[Fig.1] a) HREM image of Bi2Sr2CaCu2O8+x high temperature superconductor,
                T$_{c}=$85K.
         The insert down left is its optical (laser) diffraction pattern.

         b) Fast Fourier Transform (FFT) of  Fig.1a. Most of inelastically
            scattered electrons
         are inside two first Brillouin zones. Inserts up right and down
         left show the  modes of
         analysis of  Fig.1a when using inelastically scattered electrons
         (see text).

\item[Fig.2] a) Photodensitometric trace of Fig.1a along [200] direction.
               Similar trace is obtained
               along [020] direction. This traces, digitalized, are used in the
               R/S analysis.

         b) FFT of the trace shown in a). c)FFT of the trace in [020]
           direction. The diffuse
           backgrounds inside (200) and (020) are analysed in present paper.

\item[Fig.3] a) Filtered image of Fig.1a. The Bragg reflections used in
                the ProFFT programm are shown nonmasked in the insert up
                right. This figure shows up "clear" crystal lattice.

         b) Double filtered image of the same detail of the crystal shown
            in Fig.1a. Fine mosaic
         substructure is now better emphasized. The appearance of faint
            reflections of the (110)
         type is not investigated here.

\item[Fig.4] a) Fig.1a highly computer magnified and superimposed on the crystal
                grid with a=b=5.4 A.U. to show that the atom columns lie
                on the crystal nodes.

             b) This is better seen in this, doubly filtered,
                high magnified detail of  Fig.3b. White arrows left
                are added as the position control mark

\item[Fig.5] a) The trace along [200] direction, containing 4096 measured points
 of the densitogram, the part of which is shown in Fig.2a.
Similar traces are obtained for all measurements analyzed in the present
paper.

b) The R/S analysis of the trace in a).

c) The "512 points" trace.

d) The R/S analyzis of the trace in c).

e) Randomized "4096" trace.

f) The R/S analyzis of randomized trace in e), showing no ordering
(H close to 0.5.).

\item[Fig.6] a) The ProFFT analysis of  diffusely scattered electrons in
                the 1st and 2nd B.z. in the
                [100] direction. Absorption and re-emission regular bands
                of inelastically scattered
                electrons can be easily recognized. b)The same argumentation
holds for [010]. Dark and white bands are nearly equidistant and of the
order of atom spacings in the crystal.

\item[Fig.7] a) Combined figure, containing the effects of the diffuse
                diffraction in 1st and 2nd B.z.,
         in the [100] and the [010] directions. See the details in the text.

b) As in a), but with the
         "narrow cross". Similarities in both figures point to the Chaotic
behaviour of the inelastically scattered electrons,
as discussed in the text.

\item[Fig.8] a) The situation for the Fermi acceleration model. b) Quantum
mechanical wave function for a quantum billiard system.  c) The ProFFT
analysis of Fig.1a using the "selection"
ring for the electrons of nearly equal total energy loss, near the
edge of the 1st Brillouin zone.
The ring takes electrons diffracted in all directions with
the energy loss so they point to the edge of the 1st Brillouin zone.
The similarity with b) is obvious.
\end{itemize}
\end{description}

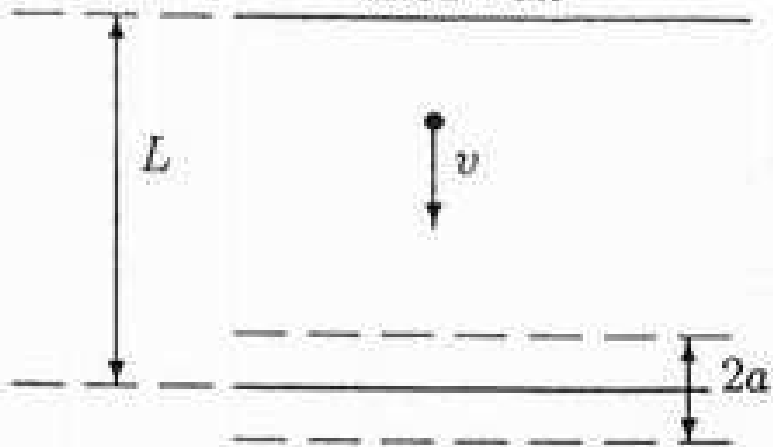
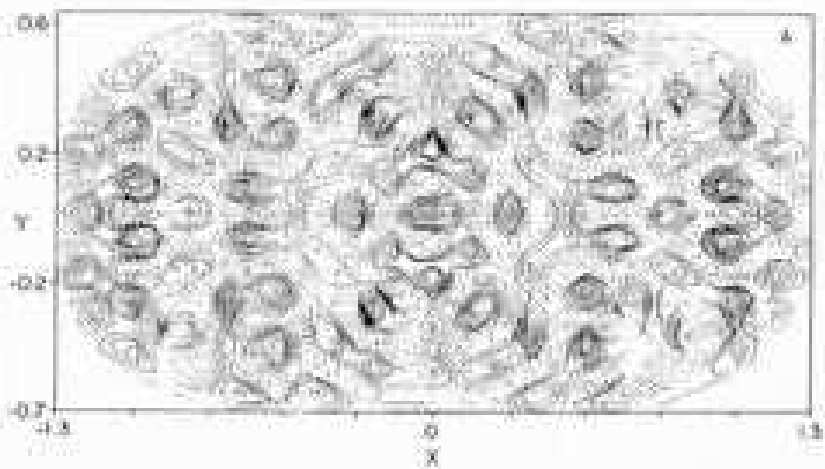
\end{document}